\documentclass[fleqn,usenatbib]{mnras}
\usepackage{newtxtext,newtxmath}
\usepackage[T1]{fontenc}
\usepackage{ae,aecompl}
\usepackage{tabularx}
\usepackage{graphicx}
\usepackage{float}
\usepackage{mathrsfs}	
\usepackage{amsmath}
\usepackage{color}
\usepackage{hyperref}
\usepackage{multirow}
\usepackage{subfigure}
\usepackage{longtable}

\usepackage{graphicx}	
\usepackage{amsmath}	
\usepackage{amssymb}	
\usepackage{csvsimple}

\title{Strong-lensing Measurement of the Mass-density Profile out to 3 Effective Radii for $z \sim 0.5$ Early-type Galaxies} 
\author[Li et al.]
{Rui Li$^{1,2,3,4}$\thanks{E-mail: lirui@ynao.ac.cn}, Yiping Shu$^{5,6}$\thanks{E-mail: yiping.shu@pmo.ac.cn}, Jiancheng Wang$^{1,2,3,4}$\thanks{E-mail: jcwang@ynao.ac.cn}    
\vspace*{0.2cm}\\
$^{1}$ Yunnan Observatories, Chinese Academy of Sciences, 396 Yangfangwang, Guandu District, Kunming, 650216, P. R. China\\ 
$^{2}$ University of Chinese Academy of Sciences, Beijing, 100049, P. R. China \\
$^{3}$Center for Astronomical Mega-Science, Chinese Academy of Sciences, 20A Datun Road, Chaoyang District, Beijing, 100012,\\ 
\ \ \ P. R. China \\ 
$^{4}$ Key Laboratory for the Structure and Evolution of Celestial Objects, Chinese Academy of Sciences, 396 Yangfangwang, Guandu \\
\ \ \ District, Kunming, 650216, P. R. China\\
$^{5}$ Purple Mountain Observatory, Chinese Academy of Sciences, 2 West Beijing Road, Nanjing, Jiangsu, 210008, China \\
$^{6}$ Institute of Astronomy, Madingley Road, Cambridge CB3 0HA, UK}

\date{Accepted 2018 June 29. Received 2018 June 09; in original form 2018 January 11}

\pubyear{2018}

\begin{document}
\label{firstpage}
\pagerange{\pageref{firstpage}--\pageref{lastpage}}
\maketitle

\begin{abstract}
We measure the total mass-density profiles out to three effective radii for a sample of 63 $z \sim 0.5$, massive early-type galaxies (ETGs) acting as strong gravitational lenses through a joint analysis of lensing and stellar dynamics. The compilation is selected from three galaxy-scale strong-lens samples including the Baryon Oscillation Spectroscopic Survey (BOSS) Emission-Line Lens Survey (BELLS), BELLS for GALaxy-Ly$\alpha$ EmitteR sYstems Survey, and Strong Lensing Legacy Survey (SL2S). Utilizing the wide source-redshift coverage (0.8--3.5) provided by these three samples, we build a statistically significant ensemble of massive ETGs for which robust mass measurements can be achieved within a broad range of Einstein radii up to three effective radii. Characterizing the three-dimensional total mass-density distribution by a power-law profile as $\rho \propto r^{-\gamma}$, we find that the average logarithmic density slope for the entire sample is $\langle\gamma\rangle=2.000_{-0.032}^{+0.033}$ ($68\%$CL) with an intrinsic scatter $\delta=0.180_{-0.028}^{+0.032}$. Further parameterizing $\langle\gamma\rangle$ as a function of redshift $z$ and ratio of Einstein radius to effective radius $R_{ein}/R_{eff}$, we find the average density distributions of these massive ETGs become steeper at larger radii and later cosmic times with magnitudes $\mathrm{d} \langle\gamma\rangle / \mathrm{d}z = -0.309_{-0.160}^{+0.166}$ and $\mathrm{d} \langle\gamma\rangle / \mathrm{d} \log_{10} \frac{R_{ein}}{R_{eff}} = 0.194_{-0.083}^{+0.092}$. 
\end{abstract}


\section{Introduction}

Believed to be the end products of the hierarchical merging scenario \citep{1993MNRAS.264..201K, 2000MNRAS.319..168C}, early-type galaxies (ETGs) play an important role in understanding the formation and evolution of galaxies. In particular, the mass-density slope in the inner region of ETGs provide useful insight into the physical processes that regulate the mass distributions. Numerical simulations suggest that dark-matter halos across a broad range of mass scales can be well described by a universal NFW profile \citep{1997ApJ...490..493N}. However, baryonic physics can significantly modify the mass distribution of an ETG, especially in the inner region, through dissipative gas-cooling processes and the supernovae (SN)/ active galactic nucleus (AGN) feedback. The former processes lead to a higher baryon densities and steepen the inner density profile \citep[e.g.,][]{2006PhRvD..74l3522G, 2010MNRAS.407..435A, 2014MNRAS.442.2641V}, while the latter processes tend to heat the gas and soften the density profile \citep[e.g.,][]{2012MNRAS.422.3081M, 2013MNRAS.433.3297D, 2014MNRAS.442.2641V}. Therefore, the inner mass-density profile of ETGs and its dependences on galaxy properties can be used to study the relative importance and efficiency of the aforementioned physical processes across different evolving stages. 

\begin{figure*}
\centering
\subfigure[]{
    \label{fig:subfig:a}
    \includegraphics[height=5.5cm,width=8cm]{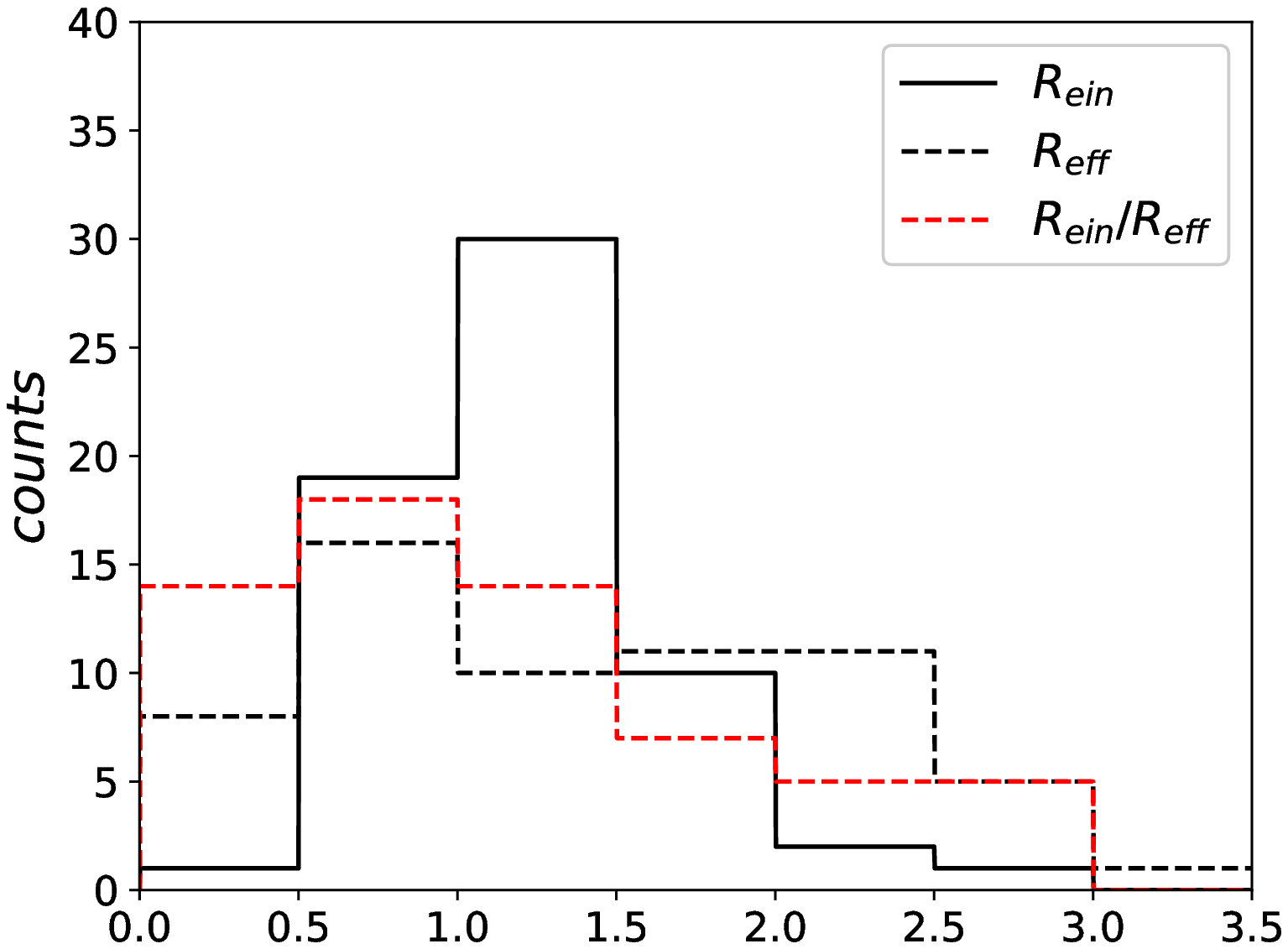}
    }
\subfigure[]{
    \label{fig:subfig:b}
    \includegraphics[height=5.5cm,width=8cm]{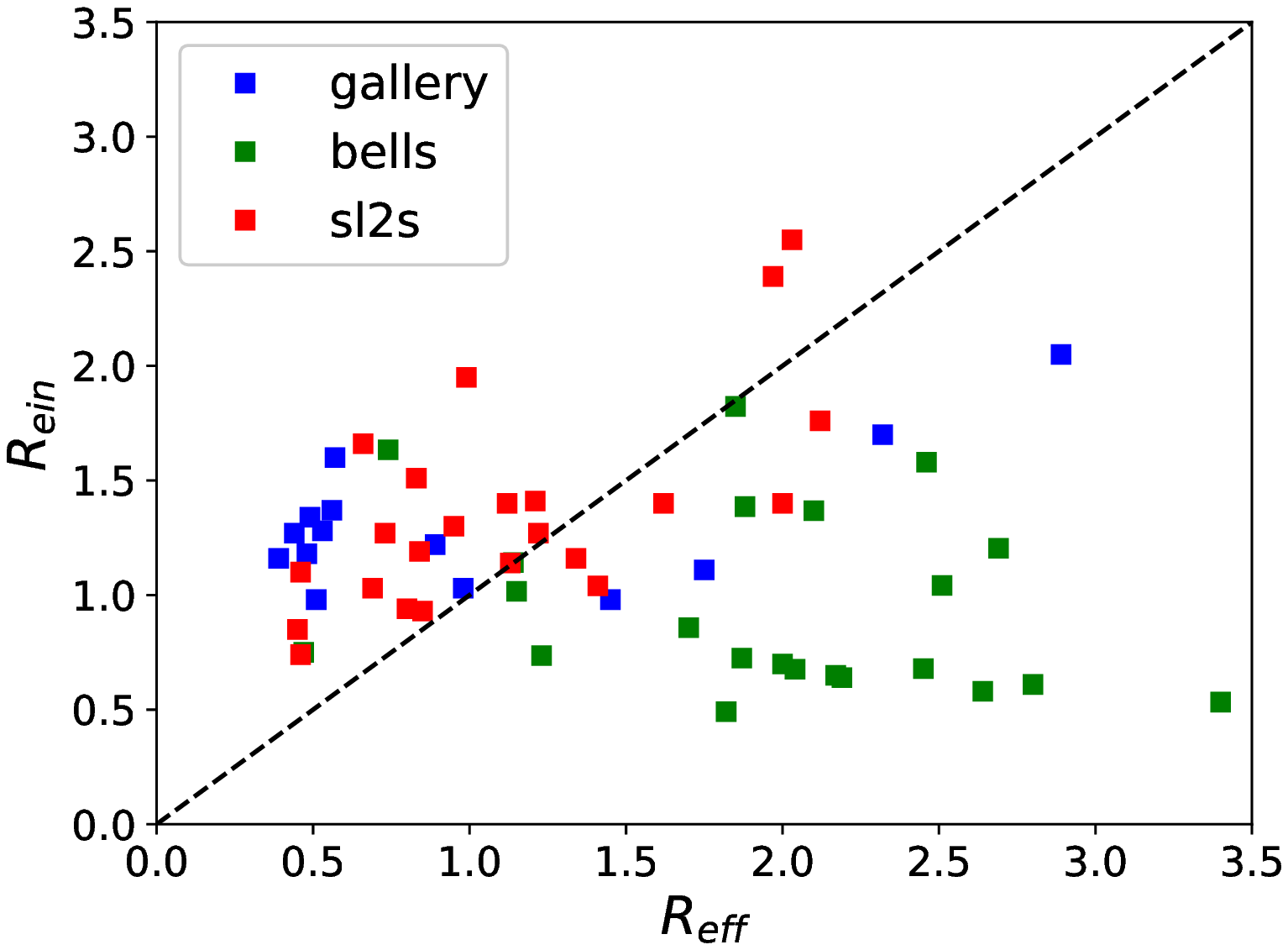}}
\caption{\label{fig:evolprob}
Fig(a) shows the distributions of the Einstein radius (black solid line), the effective radius (black dashed line), and the ratio of the Einstein radius to effective radius (red dashed line) for the lens compilation considered in this work. Fig(b) shows the comparison between the Einstein radius and the effective radius for individual lens samples.}
\end{figure*}

One effective method of measuring the inner mass distributions of ETGs is through stellar dynamical modeling. For instance, \cite{2015ApJ...804L..21C} analyzed two-dimensional stellar kinematic data for 14 local fast-rotator ETGs to infer the mass density profiles of this sample out to 4 effective radii. However, for distant ETGs ($z \gtrsim0.2$), stellar kinematic observations, especially the integral field unit (IFU) observations, become technically challenging and sometimes impossible. Alternatively, strong gravitational lensing serves as another powerful method of measuring the mass distributions of distant ETGs. Assuming a two-parameter power-law mass distribution model, the projected total mass constraint within the Einstein radius provided by the strong-lensing data and the stellar velocity dispersion measured within a fiber or slit can be used to determine the average mass-density slope of the lens galaxy in its central region by solving the spherical Jeans equations. This joint analysis of strong lensing and stellar dynamics has been applied to a sample of lens galaxies, which leads to better understanding of the mass distribution and its evolution for massive ETGs. For instance, \cite{2006ApJ...649..599K} found that the average inner mass-density profile of massive ETGs can be well approximated by an isothermal profile (i.e. $\gamma=2$) based on 15 lenses selected from the Sloan Lens ACS Survey \citep[SLACS,][]{2008ApJ...682..964B}. \cite{2010ApJ...724..511A} and \cite{2013ApJ...777...98S} suggested that the average logarithmic density slope correlates with the central surface mass density in the sense that denser galaxies have steeper slopes. \citet{2012ApJ...757...82B} pointed out that inner density slope of massive ETGs evolves with galaxy redshift. \cite{2015ApJ...803...71S} found that the inner density slope and dark matter fraction are strongly correlated with galaxy mass/velocity dispersion. 

Another important question is whether the mass-density slope evolve along radius. If such an
evolution indeed exists, it could affect the interpretations of some results based on strong lenses,
in particular the redshift evolution of the density slope in massive galaxies. \cite{2006ApJ...649..599K} found little correlation between the mass-density slope and the normalized radius $R_{ein}/R_{eff}$, but their results were limited to within one effective radius and lower redshifts ($z<0.3$). Although \cite{2013ApJ...777...98S} extended this analysis to 3 effective radii and higher redshifts ($0.3<z<0.7$), the results were not statistically significant because only two of the 23 lens galaxies used in their work have $R_{ein}/R_{eff}$ larger than 2.

In this work, we combine BELLS and SL2S lenses with the recently discovered BELLS GALLERY lenses, which contains 7 lens galaxies with $R_{ein}/R_{eff} > 2$, to examine the radial dependence of the mass-density slope of massive ETGs at high redshifts ($0.3<z<0.7$). Throughout the paper, $R$ represents the two-dimensional radial coordinate and $r$ represents the three-dimensional radial coordinate. We adopt a fiducial cosmological model with $\rm \Omega_m = 0.274$, $\rm \Omega_{\Lambda} = 0.726$, and $H_0 \rm = 70\,km\,s^{-1}\,Mpc^{-1}$.

\section{Lens systems and the data}
Strong gravitational lenses can only provide a robust measurement of the total mass within the Einstein radii. To infer the mass distribution, we assume the total-mass density distribution of a lens galaxy can be well described by a power-law profile as suggested by a variety of studies \citep{Koopmans03, 2006ApJ...649..599K, 2009ApJ...703L..51K, 2010ApJ...724..511A, 2012ApJ...757...82B, 2013ApJ...777...98S, 2015ApJ...803...71S}, and use the stellar velocity dispersion as an extra constraint for the logarithmic slope $\gamma$. We then build an ETG-lens compilation with a wide range of Einstein radii $R_{ein}$ and study the radial dependence of $\gamma$ for entire ETG-lens population within a hierarchical Bayesian framework. This is achieved by selecting strong-lens systems with similar lens masses and redshifts but a wide range of source redshifts because the Einstein radii is determined by the lens' mass, distance to the observer, and distance to the source as $R_{ein} \propto [M(<R_{ein}) (1-D_d/D_s)]^{1/2}$, where $D_d$ and $D_s$ are the comoving distances to the lens and source respectively. Given a similar lens mass and redshift, a wide range of source redshifts corresponds to a wide range of Einstein radii.

The compilation of the 63 strong-lens systems used in this work is built from the BELLS, BELLS GALLERY, and part of SL2S samples. BELLS and BELLS GALLERY strong-lens systems were selected from the BOSS spectroscopic database based on detections of emission lines in the BOSS galaxy spectra that are identified to come from redshifts higher than the BOSS galaxies \citep{2006aglu.confE...3B, 2006ApJ...638..703B}. High-resolution \textsl{HST} imaging data were acquired to confirm the lensing nature. SL2S strong-lens systems were identified photometrically from the CFHT Legacy Survey by the presence of lensing-like features around galaxies and confirmed by imaging (\textsl{HST} or \textsl{CFHT}) and spectroscopic data. In total, there are 67 strong-lens systems in the three samples with 25 from BELLS \citep{2012ApJ...744...41B}, 17 from BELLS GALLERY \citep{2016ApJ...833..264S}, and 25 from SL2S \citep{2013ApJ...777...97S, 2013ApJ...777...98S}. The lens redshifts of the three lens samples are within a similar range of 0.3--0.65, but the source redshifts cover a wide range from 0.8 to 3.5. Because this work aims to investigate the mass-density profile of massive ETGs, we discard two BELLS GALLERY systems with multiple lens components and two SL2S systems with disk-like lenses. The remaining 63 strong-lens systems, each with a single massive ETG as the lens, are used in this analysis.

Figure 1a shows the distributions of the Einstein radius ($R_{ein}$, black solid line), effective radius ($R_{eff}$, black dashed line), and ratio of the Einstein radius to the effective radius ($R_{ein}/R_{eff}$, red dashed line) for the 63 lens galaxies. The $R_{ein}/R_{eff}$ ratios populate primarily between 0.5 and 1.5 with a sharp drop below 0.5 and an extended wing up to 3.5. Ten galaxies have $R_{ein}/R_{eff}$ ratios larger than 2.0. In Figure 1b, we compare the Einstein radii ($R_{ein}$) and effective radii ($R_{eff}$). We find that the Einstein radii are smaller than the effective radii for most of the BELLS and SL2S samples. But for the BELLS GALLERY samples, the Einstein radii of most systems are larger than the effective radii.

\section{Joint lensing and dynamical analysis}

Following previous work \citep[e.g., ][Bolton et al. 2012]{2004ApJ...611..739T, 2006ApJ...649..599K, 2009ApJ...703L..51K}, we assume the three-dimensional total mass-density distribution of the lens galaxies can be described by a power-law profile as 
\begin{equation}
\rho_{tot}=\rho_{0}(r)^{-\gamma},
\end{equation}
where $\gamma$ is the logarithmic slope and $\rho_{0}$ is a normalization factor.
We model the two-dimensional light distributions of lens galaxies using a S\'{e}rsic profile as 
\begin{equation}
I(R)=I_e \exp \{-b_n[(\frac{R}{R_{eff}})^\frac{1}{n}-1]\},
\end{equation}
where $n$ is the S\'{e}rsic index, $b_n$ is given by $b_n \approx 1.9992n-0.3271$ for $1 <n<10$, and $R_{eff}$ is the effective radius. We de-project two-dimensional light distributions to infer the three-dimensional light distributions by solving the Abel integral equation. 
The spherical Jeans Equation can be written as \citep{1987gady.book.....B}
\begin{equation}
\frac{1}{\nu} \frac{d(\nu \bar{v_r ^2})}{dr}+2 \frac{\beta \bar{v_r ^2}}{r}=-\frac{GM(<r)}{r^2},
\end{equation}
where $v_\theta$ and $v_r$ are the tangential and radial components of velocity dispersion vector. $\nu$ is the number density of the stars in the galaxy, which is assumed to be proportional to the three-dimensional light distributions. $\beta=1-\frac{v_\theta^2}{v_r^2}$ is the velocity dispersion anisotropy parameter \citep{1979PAZh....5...77O, 1985AJ.....90.1027M}, which is set to be 0 in our analysis. $M(<r)$ is the total mass inside a sphere with radius $r$. 

For each lens galaxy, we use the total enclosed mass within the Einstein radius provided by strong-lensing data to determine its mass-density normalization factor $\rho_{0}$. The three-dimensional velocity dispersion profile, $\sigma(r, \gamma)$, is determined by solving Equation (3). Finally we project $\sigma(r, \gamma)$ to get the two-dimensional velocity dispersion profile, $\sigma_p(R, \gamma)$, for each lens galaxy.
In order to compare with observations, which measure the luminosity-weighted velocity dispersion within an aperture, we further convolve $\sigma_p(R, \gamma)$ with the two-dimensional light distribution $I(R)$ and an aperture weighting function $\omega(R)$ to get a predicted velocity dispersion, $\sigma_{pred}$ as 
\begin{equation}
\sigma_{pred}^2(\gamma)=\frac{\int_0^\infty dR \; R \; \omega(R) \;I(R) \; \sigma_p^2(R, \gamma)}{\int_0^\infty dR \; R \; \omega(R) \; I(R)}.
\end{equation}
We adopt the functional form in \cite{2010ApJ...708..750S} for the weighting function. An aperture radius of 1$^{\prime \prime}$ and the mean seeing value 1.5855 of the 40 BELLS+BELLS GALLERY lens galaxies are used in $\omega(R)$.  

Instead of directly comparing $\sigma_{pred}(\gamma)$ to the reported velocity dispersions $\sigma_{BOSS}$, which generally have large uncertainties because they are determined from low signal-to-noise ratio (SNR) spectroscopic data, we use $\chi^2(\sigma | d_i)$, which is the $\chi^2$ curve as a function of trial velocity dispersion $\sigma$, to infer $\gamma$ for each lens galaxy. As explained in Shu et al. (2012) and Bolton et al. (2012), the $\chi^2(\sigma | d_i)$ is obtained by fitting the same spectroscopic data, $d_i$, using fewer eigenspectra to avoid over-fitting. Redshift error is also marginalized in this process. We minimize $\chi^2(\gamma | d_i)$, which is converted from $\chi^2(\sigma | d_i)$ by interpolations, to determine the best-fit $\gamma$ value for each lens galaxy. 

The logarithmic total mass-density profile slope $\gamma$ for BELLS and BELLS GALLERY lens galaxies are obtained by this analysis, and listed in the last column of Table~\ref{tb:tb}. The logarithmic total mass-density slope of SL2S lens galaxies as well as their  uncertainties are directly taken from \cite{2013ApJ...777...98S}, who performed a similar joint analysis.

\section{The average total mass-density slope and its radial dependence}

In this section, we perform a hierarchical Bayesian analysis to study the mean total mass-density slope of this lens galaxy sample and its dependences on other galaxy properties, especially the ratio of the Einstein radius to the effective radius. Here we assume the 63 lens galaxies can be treated as a single population. 

We parameterize the probability density function (PDF) of $\gamma$ as a Gaussian function
\begin{equation}
p(\gamma | \langle\gamma\rangle, \delta_{\gamma})= {1 \over{\sqrt{2 \pi} \delta_{\gamma}}} \exp \left\{ - {{(\gamma - \langle\gamma\rangle )^2} \over {2 \delta_{\gamma}^2}} \right\},
\end{equation}
where $\langle\gamma\rangle$ and $\delta_{\gamma}$ are the two hyper-parameters that represent the mean and intrinsic scatter of the mass-density slope. 
The likelihood function of $\langle\gamma\rangle$ and $\delta_{\gamma}$, $\mathscr{L} (\langle\gamma\rangle,\delta_{\gamma} | \vec{d})$, can be written as 
\begin{align}
\mathscr{L} (\langle\gamma\rangle,\delta_{\gamma} | \vec{d}) &= p(\vec{d} | \langle\gamma\rangle,\delta_{\gamma}) = \prod_{i=1}^{60} p({d_i} | \langle\gamma\rangle,\delta_{\gamma}) \nonumber \\
&= \prod_{i=1}^{60} \int \mathrm{d} \gamma \, p({d_i} | \gamma) p(\gamma | \langle\gamma\rangle,\delta_{\gamma}),
\end{align}
in which $p({d_i} | \gamma)$ is simply proportional to $\exp[-\chi^2(\gamma | d_i)/2]$. Notice that, for SL2S lens galaxies, we have no $-\chi^2(\sigma | d_i)$ curve. Therefore, we use the given values of $\gamma$ and their uncertainties in \cite{2013ApJ...777...98S} to build a simple $-\chi^2(\gamma)$ curve for our study. Then the posterior PDF of $\langle\gamma\rangle$ and $\delta_{\gamma}$ is obtained from the likelihood function through the Baye's theorem as 
\begin{equation}
p(\langle\gamma\rangle,\delta_{\gamma}|\vec{d}) \propto p(\vec{d} | \langle\gamma\rangle, \delta_{\gamma}) \, p(\langle\gamma\rangle, \delta_{\gamma}),
\end{equation}
where $p(\langle\gamma\rangle, \delta_{\gamma})$ is the prior. For simplicity, we adopt an uniform prior over reasonable ranges of $\langle\gamma\rangle$ and $\delta_{\gamma}$. 

The posterior PDF contours of $\langle\gamma\rangle$ and $\delta_{\gamma}$, constructed from a Markov chain Monte Carlo method, is shown in Figure 2. We find that the total mass-density slope of this lens galaxy population is $\langle\gamma\rangle=2.000_{-0.032}^{+0.033}$, and its intrinsic scatter is $\delta_{\gamma}=0.180_{-0.028}^{+0.032}$. It suggests that the mass-density profile in the central region of this lens galaxy population is very close to an isothermal distribution, consistent with previous findings \citep{2009ApJ...703L..51K, 2010ApJ...724..511A, 2012ApJ...757...82B, 2013ApJ...777...98S, 2017ApJ...840...34S, 2018arXiv180206629M}.

\begin{figure}
\centering
\includegraphics[height=0.35\textwidth, width=0.45\textwidth]{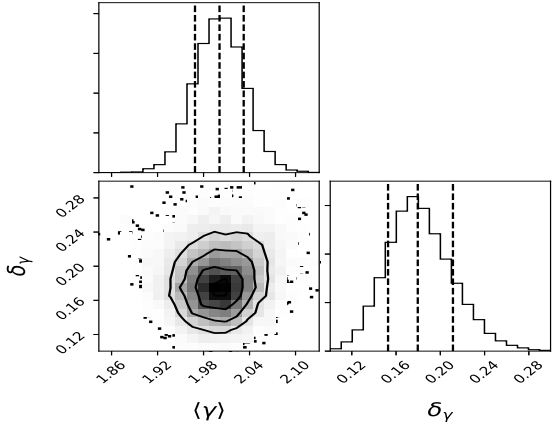}
\caption{\label{fig:illustration}
The posterior probability contours and credible regions for the mean total mass-density slope $\langle\gamma\rangle$ and the intrinsic scatter $\delta_{\gamma}$. The regions between the left and right dashed lines indicate 68\% confidence level. The middle dashed lines are the mean values of the parameters.}
\end{figure}

We now examine the radial dependence of $\gamma$ in this galaxy population. To do this, we normalize the Einstein radius by the effective radius for each lens galaxy, and introduce one more hyper-parameter, $\alpha$, so that the PDF of $\gamma$ is now 

\begin{align}
p(\gamma | \gamma_0, \alpha, \delta_{\gamma})
&= {1 \over{\sqrt{2 \pi} \delta_{\gamma}}}\nonumber \\
&\times \exp \left\{ - {{ \left\{\gamma - [\gamma_0 + \alpha (log_{10}(\frac{R_{ein}}{R_{eff}})+0.05)]\right\}^2} \over {2 \delta_{\gamma}^2}} \right\},
\label{paramdist}
\end{align}
where $\gamma_0$ is the logarithmic slope at $log_{10}(R_{ein}/R_{eff})=-0.05$, which is the mean $log_{10}(R_{ein}/R_{eff})$ value for the full lens-galaxy population, $\alpha$ quantifies the radial dependence of $\gamma$, and $\delta_{\gamma}$ is the intrinsic scatter. Finally we get $\gamma_0=1.983_{-0.033}^{+0.033}$, $\alpha=0.170_{-0.118}^{+0.121}$, and $\delta_{\gamma}=0.176_{-0.028}^{+0.031}$. The fitting result is shown in Figure 3. We find that, the slope $\gamma$ has a very slight increasing trend along the radii within 3 effective radii for a fixed galaxy.
\begin{figure}
\centering
\includegraphics[height=0.30\textwidth, width=0.45\textwidth]{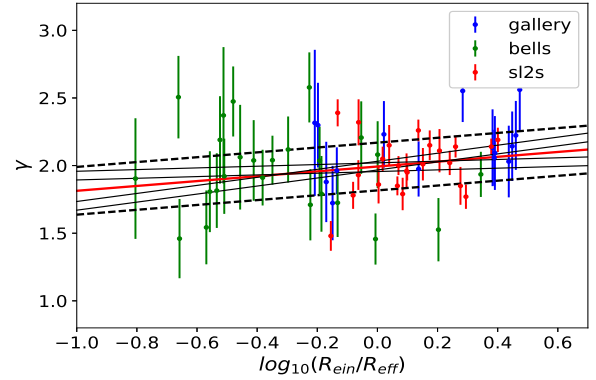}
\caption{\label{fig:illustration}
The red solid line shows the best fitting relation, the gray lines indicate the $1\sigma$ errors of $\gamma_0$ and $\gamma_R$, the dashed lines indicate the intrinsic scatter. The data points are the minimum-$\chi^2$ value for the logarithmic total mass-density profile slope $\gamma$ for the BELLS (green), BELLS GALLERY (blue) and SL2S (red) lenses. The error bars indicate the $1\sigma$ error of $\gamma$ for each galaxy.}
\end{figure}

\begin{figure*}
\centering
\includegraphics[height=12cm,width=17cm]{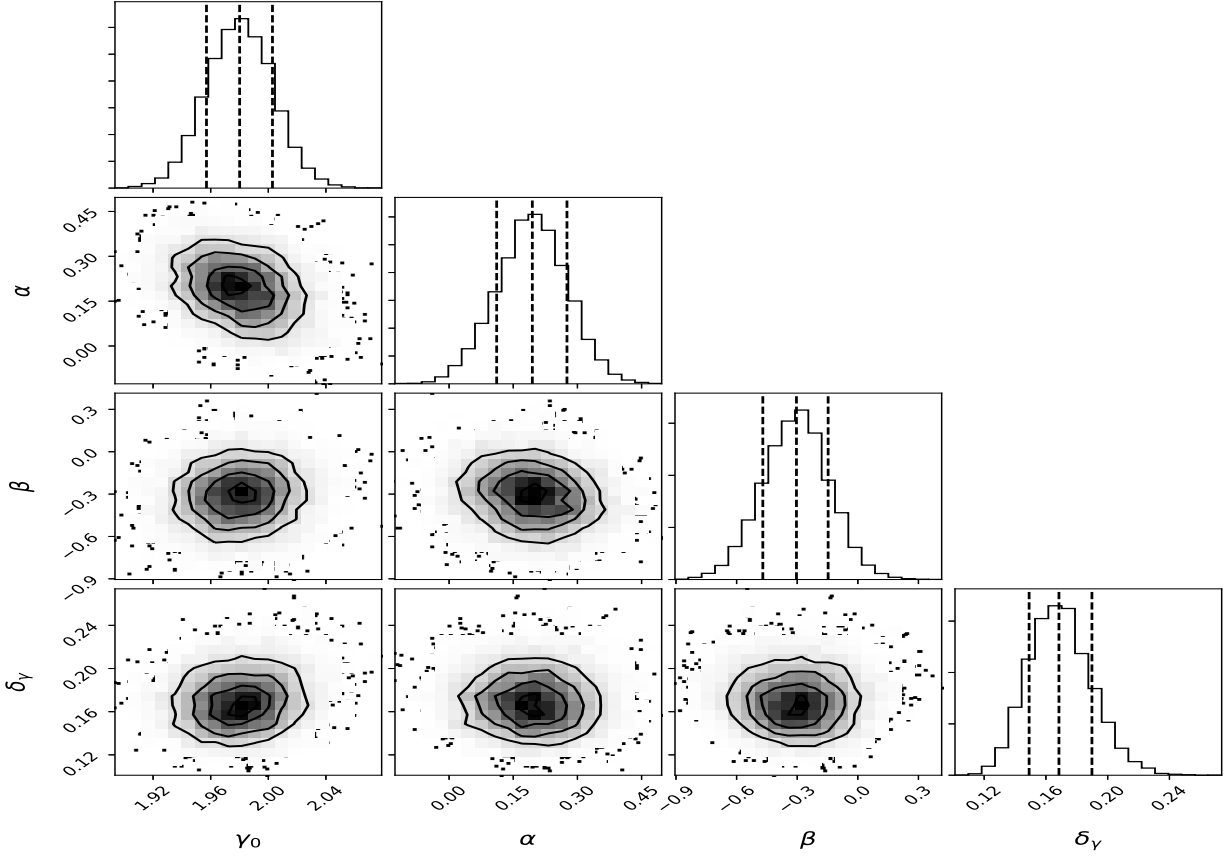}
\caption{\label{fig:evolprob}
The posterior probability contours and credible regions for $\gamma_{0}$, $\alpha$, $\beta$, and $\delta_{\gamma}$. The regions between the left and right dashed lines indicate 68\% confidence level. The middle dashed lines are the mean values of the parameters.}
\end{figure*}

\section{discussion}
The time evolution of slope $\gamma$ that galaxies with lower redshift would have steeper mass-density profile, is clearly visible in previous work\citep{2012ApJ...757...82B}. For our samples, we also examine the slope evolution with redshift. Assuming this relation is linear, we write the PDF as
\begin{align}
p(\gamma | \gamma_0, \beta, \delta_{\gamma})
&= {1 \over{\sqrt{2 \pi} \delta_{\gamma}}}\nonumber \\
&\times \exp \left\{ - {{ \left\{\gamma - [\gamma_0 + \beta (z-0.52)]\right\}^2} \over {2 \delta_{\gamma}^2}} \right\},
\label{paramdist}
\end{align}
where $\gamma_0$ is the logarithmic slope at $z=0.52$, which is the mean redshift value for all the samples, $\beta$ quantifies the redshift dependence of $\gamma$. The result is $\beta=-0.255_{-0.226}^{+0.232}$, indicating that the galaxies at higher redshift have shallower mass density profile, which are similar with previous work \citep{2010ApJ...724..511A, 2012ApJ...757...82B}.

In our research, by normalizing the Einstein radius by the effective radius for each lens galaxy, we find $\gamma$ increase along radius, which is a fundamental dependence of the slope on structural properties of ETGs. However, we notice that, with redshift increasing, the ratio of $R_{ein}/R_{eff}$ would increase for pure geometrical reasons. Therefore, the existence of the time evolution of $\gamma$ may affect the inference of the radius evolution and vice versa. So, in order to make an comprehensive analysis, we assume the slope $\gamma$ evolutes with $R_{ein}/R_{eff}$ and the redshift of the lenses at the same time as
\begin{equation}
p(\gamma | \gamma_0, \alpha, \beta, \delta_{\gamma})= {1 \over{\sqrt{2 \pi} \delta_{\gamma}}} \exp \left\{ - {{ (\gamma - \langle\gamma\rangle)^2} \over {2 \delta_{\gamma}^2}} \right\}
\end{equation}
where
\begin{equation}
\langle\gamma\rangle =\gamma_0+\alpha(log_{10}(\frac{R_{ein}}{R_{eff}}) + 0.05)+\beta(z-0.52).
\end{equation}
$\gamma_0$ is the value of the slope when $log_{10}(R_{ein}/R_{eff})=-0.05$, and $z=0.52$, which are the mean values of $log_{10}(R_{ein}/R_{eff})$ and redshift $z$, respectively. $\alpha$, $\beta$ are the evolution factors. The posterior probability contours and the credible regions for all these parameters are shown in Figure 4. We get the parameters as follows
\begin{eqnarray*}
\gamma_0 = 1.981_{-0.024}^{+0.024},  \\
\alpha =  0.194_{-0.083}^{+0.092},  \\
\beta = -0.309_{-0.160}^{+0.166},\\
\delta_{\gamma} = 0.168_{-0.017}^{+0.021}.\\
\end{eqnarray*}
Besides the time evolution, we find the slope $\gamma$ still has a increasing trend along radius, implying that the total mass-density slope $\gamma$ would become steeper with increasing radius for a fixed galaxy.

Unlike previous work (e.g., \cite{2006ApJ...649..599K, 2009ApJ...703L..51K, 2011ApJ...727...96R, 2013ApJ...777...98S}, we find $\gamma$ has a slight increasing trend along radius within 3 effective radius. This finding can be mostly attributed to the larger $R_{ein}/R_{eff}$ of BELLS GALLERY samples with higher source redshift. One possible explanation for the evolution along radius is the different strength of SN/AGN feedback processes at different radius. The SN/AGN feedback process of ETGs may toward weaker with radius increasing, which could lead the density profile at larger radius be steeper than that at smaller radius. Actually, \cite{2017MNRAS.469.1824X} have found a mild increase of $\gamma$ with increasing $R_{ein}/R_{eff}$ using numerical simulation depending on galactic wind and AGN feedback. In their work, they used the Illustris simulation project \citep{2014MNRAS.445..175G, 2014MNRAS.444.1518V, 2014Natur.509..177V, 2015A&C....13...12N, 2015MNRAS.452..575S} to study the mass-density slope of the inner regions of ETGs and find the simulation predicted higher central dark matter fractions, which would suppress the dominating role of baryons and thus lead to shallower total mass-density profile at smaller radius. Our observational result for the density profile is in accordance with their predictions.

\section{Conclusions}
In this paper, we study the total mass-density profile out to three effective radii for a sample of 63
 intermediate-redshift early-type galaxies (ETGs). The sample is compiled from three galaxy-scale strong-lens surveys, including 25 galaxies from the BELLS,  15 galaxies from the BELLS GALLERY, and 23 galaxies from the SL2S. Assuming a power-law total-mass density profile of $\rho_{tot}=\rho_{0} r^{-\gamma}$ for the lens galaxies, we investigate the evolution of the slope $\gamma$ out to 3 effective radii with a hierarchical Bayesian method by combining the strong lensing data and dynamical constraints. In our study, we include more higher redshift galaxies in the sample comparing with the SLACS. We obtain the following conclusions:

\begin{enumerate}
\item The average logarithmic density slope of our samples is $\langle\gamma\rangle=2.000_{-0.032}^{+0.033}$, with an intrinsic scatter of $\delta=0.180_{-0.028}^{+0.032}$. The total mass-density profile is very close to the isothermal distribution, similar with previous work.

\item Assuming a linear relation between $\gamma$ and $log_{10}(R_{ein}/R_{eff})$, we obtain that the evolution factor is $\alpha=0.170_{-0.118}^{+0.121}$, indicating that the total mass-density slope has a sight rise along radius for a fixed galaxy. When consider the slope $\gamma$ evolves with $R_{ein}/R_{eff}$ and the redshift at the same time, the increasing trend still exists. We conclude that the total mass-density slope of ETGs would increase along radius within 3 effective radii for a fixed galaxy, which is in accordance with the numerical simulation depending on galactic wind and AGN feedback \citep{2017MNRAS.469.1824X}.
\end{enumerate}

We acknowledge the financial support from the National Natural Science Foundation of China 11573060 and 11661161010. Y.S. has been supported by the National Natural Science Foundation of China 
(No. 11603032 and 11333008), the 973 program (No. 2015CB857003), and 
the Royal Society - K.C. Wong International Fellowship (NF170995).

\appendix
\section{List of strong gravitational lenses}\label{sec:app}

Table~\ref{tb:tb} shows a list of all 63 strong gravitational lens
systems used in this paper.

\onecolumn
\setlongtables
\begin{longtable}{c c c c c c c c c c}
\caption{Column 1 is the lensing systems name, the first 15 systems are from the BELLS GALLERY, the next 25 are from the BELLS and the rest are from the SL2S. Columns 2 and 3 are the redshifts of the foreground lenses and the background galaxies inferred from the BOSS spectrum. Column 4 is the velocity dispersion. Column 5 and Column 6 are the Einstein radius and the effective radius. Column 7 is the Einstein mass inside the Einstein radius. Column 8 is the effective slope $\gamma$ assuming a power-law mass density distribution. Two BELLS GALLERY systems with multiple lens components and two SL2S systems with disk-like lenses are not shown here. 
\label{tb:tb}}\\
\hline \hline
Lens Name & $z_{L}$ & $z_{S}$ & $\sigma_{\rm BOSS}$ & $R_{\rm ein}$ & $R_{\rm eff}$ & $M_{\rm ein}$ & Slope $\gamma$\\
 & & & (km s$^{-1}$) & (arcsec) &  (arcsec) &($10^{11} M_\odot$) & ($10^{11} M_\odot$) & \\
(1) & (2) & (3) & (4) & (5) & (6) & (7) & (8) \\
SDSSJ0029$+$2544 & 0.5869 & 2.4504 & 241$\pm$45 & 1.34 & 0.49 & 4.82 &  2.03$\pm$0.27\\
\hline
SDSSJ0201$+$3228 & 0.3957 & 2.8209 & 256$\pm$20 & 1.70 & 2.32 & 5.21 &  1.96$\pm$0.17\\
\hline
SDSSJ0237$-$0641 & 0.4859 & 2.2491 & 290$\pm$89 & 0.65 & 1.05 & 0.97 &  2.32$\pm$0.54\\
\hline
SDSSJ0742$+$3341 & 0.4936 & 2.3633 & 218$\pm$28 & 1.22 & 0.89 & 3.41 &  1.98$\pm$0.20\\
\hline
SDSSJ0755$+$3445 & 0.7224 & 2.6347 & 272$\pm$52 & 2.05 & 2.89 & 13.52& 1.72$\pm$0.28\\
\hline
SDSSJ0856$+$2010 & 0.5074 & 2.2335 & 334$\pm$54 & 0.98 & 0.51 & 2.30 &  2.55$\pm$0.23\\
\hline
SDSSJ0918$+$5104 & 0.5811 & 2.4030 & 298$\pm$49 & 1.60 & 0.57 & 6.85 &  2.14$\pm$0.26\\
\hline
SDSSJ1110$+$2808 & 0.6073 & 2.3999 & 191$\pm$39 & 0.98 & 1.45 & 2.69 &  1.88$\pm$0.30\\
\hline
SDSSJ1110$+$3649 & 0.7330 & 2.5024 & 531$\pm$165& 1.16 & 0.39 & 4.48 &  2.56$\pm$0.31\\
\hline
SDSSJ1116$+$0915 & 0.5501 & 2.4536 & 274$\pm$55 & 1.03 & 0.98 & 2.68 &  2.23$\pm$0.25\\
\hline
SDSSJ1141$+$2216 & 0.5858 & 2.7624 & 285$\pm$44 & 1.27 & 0.44 & 4.18 &  2.22$\pm$0.26\\
\hline
SDSSJ1201$+$4743 & 0.5628 & 2.1258 & 239$\pm$43 & 1.18 & 0.48 & 3.76 &  2.09$\pm$0.27\\
\hline
SDSSJ1226$+$5457 & 0.4980 & 2.7322 & 248$\pm$26 & 1.37 & 0.56 & 4.20 &  2.06$\pm$0.21\\
\hline
SDSSJ2228$-$1205 & 0.5305 & 2.8324 & 255$\pm$50 & 1.28 & 0.53 & 3.85 &  2.13$\pm$0.28\\
\hline
SDSSJ2342$-$0120 & 0.5270 & 2.2649 & 274$\pm$43 & 1.11 & 1.75 & 3.05 &  2.30$\pm$0.31\\
\hline
SDSSJ0151$+$0049 & 0.5171 & 1.3636 & 219$\pm$39 & 0.68 & 2.04 & 1.37 &  2.47$\pm$0.26\\
\hline
SDSSJ0747$+$5055 & 0.4384 & 0.8983 & 328$\pm$60 & 0.75 & 1.27 & 1.83 & 2.58$\pm$0.26\\
\hline
SDSSJ0747$+$4448 & 0.4366 & 0.8966 & 281$\pm$52 & 0.61 & 2.80 & 1.19 &  2.51$\pm$0.31\\
\hline
SDSSJ0801$+$4727 & 0.4831 & 1.5181 & 98$\pm$24  & 0.49 & 1.82 & 0.63 &  1.54$\pm$0.27\\
\hline
SDSSJ0830$+$5116 & 0.5301 & 1.3317 & 268$\pm$36 & 1.14 & 1.14 & 4.10 & 2.08$\pm$0.25\\
\hline
SDSSJ0944$-$0147 & 0.5390 & 1.1785 & 204$\pm$34 & 0.73 & 1.87 & 1.85 &  2.04$\pm$0.30\\
\hline
SDSSJ1159$-$0007 & 0.5793 & 1.3457 & 165$\pm$41 & 0.68 & 2.22 & 1.64 &  2.37$\pm$0.51\\
\hline
SDSSJ1215$+$0047 & 0.6423 & 1.2970 & 262$\pm$45 & 1.37 & 2.10 & 7.95 & 1.79$\pm$0.28\\
\hline
SDSSJ1221$+$3806 & 0.5348 & 1.2844 & 187$\pm$48 & 0.70 & 2.00 & 1.59 &  2.06$\pm$0.39\\
\hline
SDSSJ1234$-$0241 & 0.4900 & 1.0159 & 122$\pm$31 & 0.53 & 3.40 & 0.98 &  1.90$\pm$0.45\\
\hline
SDSSJ1318$-$0104 & 0.6591 & 1.3959 & 177$\pm$27 & 0.68 & 2.45 & 1.91 &  1.81$\pm$0.30\\
\hline
SDSSJ1337$+$3620 & 0.5643 & 1.1821 & 225$\pm$35 & 1.39 & 1.88 & 7.23 &  1.73$\pm$0.25\\
\hline
SDSSJ1349$+$3612 & 0.4396 & 0.8926 & 178$\pm$18 & 0.75 & 0.47 & 1.83 &  1.53$\pm$0.23\\
\hline
SDSSJ1352$+$3216 & 0.4634 & 1.0341 & 161$\pm$21 & 1.82 & 1.85 & 10.33&  1.46$\pm$0.19\\
\hline
SDSSJ1522$+$2910 & 0.5553 & 1.3108 & 166$\pm$27 & 0.74 & 1.23 & 1.83 &  1.71$\pm$0.26\\
\hline
SDSSJ1541$+$1812 & 0.5603 & 1.1133 & 174$\pm$24 & 0.64 & 2.19 & 1.61 &  1.82$\pm$0.28\\
\hline
SDSSJ1542$+$1629 & 0.3521 & 1.0233 & 210$\pm$16 & 1.04 & 2.51 & 2.32 &  1.91$\pm$0.21\\
\hline
SDSSJ1545$+$2748 & 0.5218 & 1.2886 & 250$\pm$37 & 1.21 & 3.91 & 4.60 &  1.92$\pm$0.28\\
\hline
SDSSJ1601$+$2138 & 0.5435 & 1.4461 & 207$\pm$36 & 0.86 & 1.70 & 2.27 &  2.12$\pm$0.24\\
\hline
SDSSJ1611$+$1705 & 0.4766 & 1.2109 & 109$\pm$23 & 0.58 & 2.64 & 0.97 &  1.46$\pm$0.29\\
\hline
SDSSJ1631$+$1854 & 0.4081 & 1.0863 & 272$\pm$14 & 1.63 & 0.74 & 6.70 &  1.94$\pm$0.17\\
\hline
SDSSJ1637$+$1439 & 0.3910 & 0.8744 & 208$\pm$30 & 0.65 & 2.17 & 1.16 &  2.19$\pm$0.32\\
\hline
SDSSJ2122$+$0409 & 0.6261 & 1.4517 & 324$\pm$56 & 1.58 & 2.46 & 9.28 &  2.04$\pm$0.25\\
\hline
SDSSJ2125$+$0411 & 0.3632 & 0.9777 & 247$\pm$17 & 1.20 & 2.69 & 3.31 &  2.04$\pm$0.18\\
\hline
SDSSJ2303$+$0037 & 0.4582 & 0.9363 & 274$\pm$31 & 1.02 & 1.15 & 3.43 &  2.21$\pm$0.27\\
\hline
SL2SJ0212$-$0555 & 0.750 & 2.74 & 267$\pm$17 & 1.27 & 1.22 & 5.31 & 2.05$\pm$0.09\\
\hline
SL2SJ0213$-$0743 & 0.717 & 3.48 & 287$\pm$33 & 2.39 & 1.97 & 16.77& 1.79$\pm$0.12\\
\hline
SL2SJ0214$-$0405 & 0.609 & 1.88 & 238$\pm$15 & 1.41 & 1.21 & 6.13 & 1.85$\pm$0.07\\
\hline
SL2SJ0217$-$0513 & 0.646 & 1.85 & 270$\pm$21 & 1.27 & 0.73 & 5.37 & 2.02$\pm$0.09\\
\hline
SL2SJ0219$-$0829 & 0.389 & 2.15 & 300$\pm$23 & 1.30 & 0.95 & 3.15 & 2.26$\pm$0.08\\
\hline
SL2SJ0225$-$0454 & 0.238 & 1.20 & 226$\pm$20 & 1.76 & 2.12 & 3.98 & 1.78$\pm$0.10\\
\hline
SL2SJ0226$-$0420 & 0.494 & 1.23 & 266$\pm$24 & 1.19 & 0.84 & 4.26 & 2.01$\pm$0.12\\
\hline
SL2SJ0232$-$0408 & 0.352 & 2.34 & 271$\pm$20 & 1.04 & 1.41 & 1.80 & 2.39$\pm$0.10\\
\hline
SL2SJ0848$-$0351 & 0.682 & 1.55 & 205$\pm$21 & 0.85 & 0.45 & 2.88 & 1.85$\pm$0.14\\
\hline
SL2SJ0849$-$0412 & 0.722 & 1.54 & 312$\pm$18 & 1.10 & 0.46 & 5.27 & 2.14$\pm$0.06\\
\hline
SL2SJ0849$-$0251 & 0.274 & 2.09 & 275$\pm$34 & 1.16 & 1.34 & 1.80 & 2.32$\pm$0.17\\
\hline
SL2SJ0855$-$0147 & 0.365 & 3.39 & 222$\pm$19 & 1.03 & 0.69 & 1.74 & 2.15$\pm$0.11\\
\hline
SL2SJ0904$-$0059 & 0.611 & 2.36 & 178$\pm$20 & 1.40 & 2.00 & 5.55 & 1.48$\pm$0.11\\
\hline
SL2SJ0959$+$0206 & 0.552 & 3.35 & 195$\pm$22 & 0.74 & 0.46 & 1.29 & 2.11$\pm$0.16\\
\hline
SL2SJ1359$+$5535 & 0.783 & 2.77 & 229$\pm$19 & 1.14 & 1.13 & 4.45 & 1.86$\pm$0.14\\
\hline
SL2SJ1404$+$5200 & 0.456 & 1.59 & 337$\pm$19 & 2.55 & 2.03 & 15.56& 1.95$\pm$0.06\\
\hline
SL2SJ1405$+$5243 & 0.526 & 3.01 & 291$\pm$21 & 1.51 & 0.83 & 5.25 & 2.14$\pm$0.08\\
\hline
SL2SJ1406$+$5226 & 0.716 & 1.47 & 258$\pm$14 & 0.94 & 0.80 & 3.96 & 2.00$\pm$0.07\\
\hline
SL2SJ1411$+$5651 & 0.322 & 1.42 & 220$\pm$23 & 0.93 & 0.85 & 1.47 & 2.15$\pm$0.15\\
\hline
SL2SJ1420$+$5630 & 0.483 & 3.12 & 228$\pm$19 & 1.40 & 1.62 & 4.16 & 1.93$\pm$0.11\\
\hline
SL2SJ2203$+$0205 & 0.400 & 2.15 & 218$\pm$21 & 1.95 & 0.99 & 7.28 & 1.77$\pm$0.09\\
\hline
SL2SJ2205$+$0147 & 0.476 & 2.53 & 326$\pm$30 & 1.66 & 0.66 & 6.01 & 2.19$\pm$0.09\\
\hline
SL2SJ2221$+$0115 & 0.325 & 2.35 & 224$\pm$23 & 1.40 & 1.12 & 3.03 & 1.96$\pm$0.13\\
\hline \hline
\end{longtable}

\end{document}